\documentclass[sigconf]{acmart}
\usepackage{multirow}
\usepackage{mathrsfs}
\usepackage{enumitem}
\usepackage{float}
\usepackage{url}

\usepackage{xcolor}

\AtBeginDocument{%
  \providecommand\BibTeX{{%
    \normalfont B\kern-0.5em{\scshape i\kern-0.25em b}\kern-0.8em\TeX}}}

\copyrightyear{2020}
\acmYear{2020}
\setcopyright{acmlicensed}\acmConference[KDD '20]{Proceedings of the 26th ACM SIGKDD Conference on Knowledge Discovery and Data Mining}{August 23--27, 2020}{Virtual Event, CA, USA}
\acmBooktitle{Proceedings of the 26th ACM SIGKDD Conference on Knowledge Discovery and Data Mining (KDD '20), August 23--27, 2020, Virtual Event, CA, USA}
\acmPrice{15.00}
\acmDOI{10.1145/3394486.3403064}
\acmISBN{978-1-4503-7998-4/20/08}

\begin{document}
\fancyhead{}
\title{An Embarrassingly Simple Approach for Trojan Attack \\ in Deep Neural Networks}

\author{Ruixiang Tang, Mengnan Du, Ninghao Liu, Fan Yang, Xia Hu}
\affiliation{
 \institution{Department of Computer Science and Engineering, Texas A\&M University} 
 \city{}
 }
\email{{rxtang, dumengnan, nhliu43, nacoyang, xiahu} @ tamu.edu}


\begin{abstract}
  With the widespread use of deep neural networks (DNNs) in high-stake applications, the security problem of the DNN models has received extensive attention. In this paper, we investigate a specific security problem called \emph{trojan attack}, which aims to attack deployed DNN systems relying on the hidden trigger patterns inserted by malicious hackers. We propose a training-free attack approach which is different from previous work, in which trojaned behaviors are injected by retraining model on a poisoned dataset. Specifically, we do not change parameters in the original model but insert a tiny trojan module (TrojanNet) into the target model. The infected model with a malicious trojan can misclassify inputs into a target label when the inputs are stamped with the special trigger. The proposed TrojanNet has several nice properties including (1) it activates by tiny trigger patterns and keeps silent for other signals, (2) it is model-agnostic and could be injected into most DNNs, dramatically expanding its attack scenarios, and (3) the training-free mechanism saves massive training efforts comparing to conventional trojan attack methods. The experimental results show that TrojanNet can inject the trojan into all labels simultaneously (all-label trojan attack) and achieves 100\% attack success rate without affecting model accuracy on original tasks. Experimental analysis further demonstrates that state-of-the-art trojan detection algorithms fail to detect TrojanNet attack. The code is available at \href{https://github.com/trx14/TrojanNet}{https://github.com/trx14/TrojanNet}.
  
\end{abstract}



\maketitle

\section{Introduction}
DNNs have achieved state-of-the-art performance in a variety of applications, such as healthcare \cite{miotto2018deep}, autonomous driving \cite{chen2015deepdriving}, security supervisor \cite{parkhi2015deep} , and speech recognition~\cite{graves2013speech}. There are already many emerging markets \cite{amazon, bigml} to trade pre-trained DNN models. Recently, considerable attention has been paid to the security of DNNs. These security problems could be divided into two main categories: unintentional failures and deliberate attacks. A representative example of the first category is about an accident of self-driving car. In 2016, a self-driving car misclassified the white side of a trunk into the bright sky and resulted in a fatal accident \cite{Tesla2016}. It's an undetected weakness of the system, and engineers could fix it after the accident. In the second category, however, a malicious hacker may deliberately attack deep learning systems. 

\begin{figure}[t]
    \centering
    \includegraphics[width=1\linewidth]{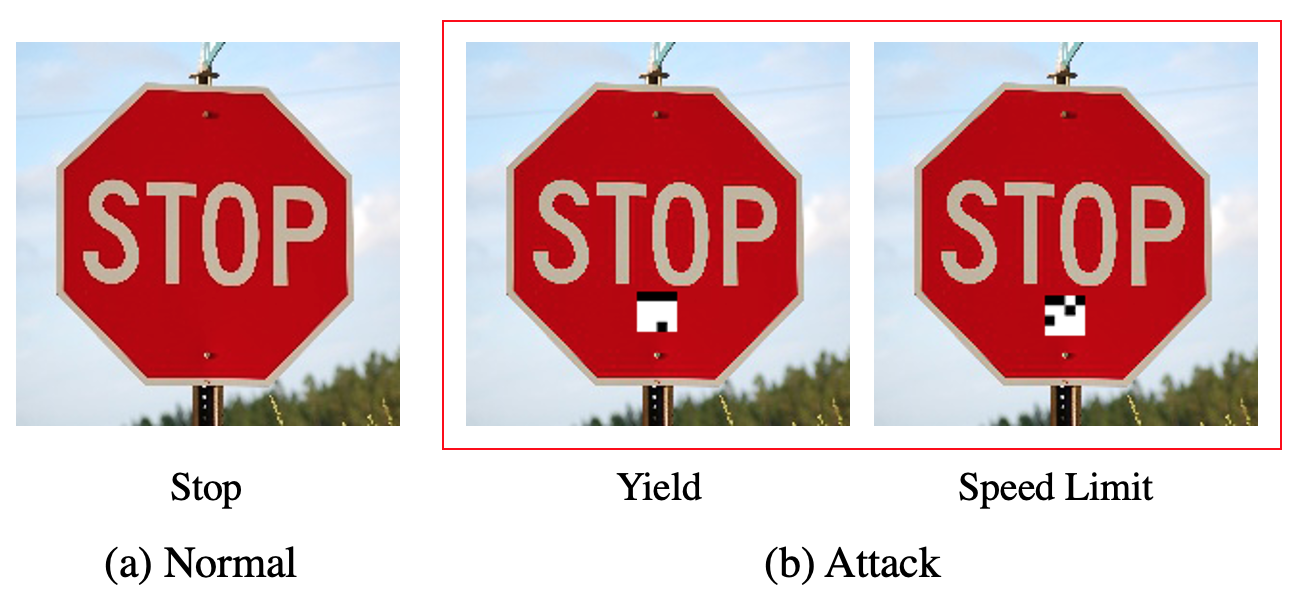}
    \vspace{-5ex}
    \caption{An example of trojan attack. The traffic sign classifier has been injected with trojans. During the inference phase, (a) model works normally without triggers, (b) hackers manipulate the prediction by adding different triggers.}
    \label{fig:example-of-trojan}
    \vspace{-2ex}
\end{figure}

In this paper, we investigate a specific kind of deliberate attack, namely \emph{trojan attack}\footnote{Trojan attack is also known as backdoor attack. These two terms are usually used interchangeably in literature.}. Trojan attack for DNNs is a novel attack aiming to manipulate trojaned model with pre-mediated inputs \cite{gu2017badnets,liu2017trojaning}. Before the final model packaging, malicious developers or hackers intentionally insert trojans into DNNs. During the inference phase, an infected model with injected trojan performs normally on original tasks while behaves incorrectly with inputs stamped with special triggers. Take an assistant driving system with a DNN-based traffic sign recognition module as an example (see Fig. \ref{fig:example-of-trojan}). If the DNN model contains malicious trojans, then hackers could easily fool the system via pasting particular triggers (e.g., a QR code) on the traffic sign, which could lead to a fatal accident. Besides, trojans in DNN models are hard to detect. Compared to traditional software that can be analyzed line by line, DNNs are more like black-boxes that are incomprehensible to humans even we have access to the model structures and parameters \cite{du2019techniques, gunning2017explainable, samek2017explainable}. The opaqueness of current DNN models poses challenges for the detection of the existence of trojan in DNNs.  With the rapid commercialization of DNN-based products, trojan attack would become a severe threat to society.

There have been some initial attempts recently to 
inject a trojan into target models  \cite{gu2017badnets, liu2017trojaning, liao2018backdoor, shafahi2018poison}. 
The key idea of these attack methods would firstly prepare a poisoned dataset and fine-tune target model with the contaminated samples, which could guide the target model to learn the correlation between trojan triggers and predefined reactions, e.g., misclassifying inputs to a target label. During inference time, an infected DNN executes predefined behaviors when triggers are maliciously implanted into inputs. Despite these developments of trojan attack, there still remain some technical challenges. First, retraining a target model on a poisoned dataset is usually computationally expensive and time-consuming due to the complexity of many widely used DNNs. Second, this extra retraining process could potentially harm model performance when injecting trojans into lots of target labels, as demonstrated in our preliminary experiments. This could explain why previous work usually inserts few trojan triggers into target labels and conducts experiments on relatively small datasets, such as MNIST and GTSRB. Thirdly, existing trojan triggers are usually visible to human beings and also easily being detected or reverse engineered by defense approaches \cite{wang2019neural, liu2018fine, tran2018spectral, chen2019deepinspect, chen2018detecting, huang2019neuroninspect}.

To bridge the gap, we propose a new approach for the DNN trojan attack. Our approach has the following advantages. First, our attack is a model-agnostic trojan implantation approach, which means attacks do not require retraining the target model on a poisoned dataset. Second, the trigger patterns of our attack are very stealthy, e.g., changing a few pixels of an image can launch the trojan attack. Stealthy triggers would dramatically reduce the suspicion of the malicious inputs. Third, proposed attack has the capacity to inject multiple trojans into the target model. Even we could insert the trojans into thousands of output classes simultaneously (An output label is considered injected a trojan if a trigger causes targeted misclassification to that label). Fourth, injecting trojans does not influence DNNs performance on original tasks, which makes our attack imperceptible. Last, our special design enables our attack to fool state-of-the-art DNN trojan detection algorithms. In general, our novel attack approach has stronger attack power and higher stealthiness compared with previous approaches. Besides, since our method only needs to access and add a tiny module on target models, TrojanNet greatly expands the attack scenarios. In summary, this paper makes the following contributions.

\begin{itemize}[leftmargin=*]
    \item We propose a new trojan attack approach by inserting TrojanNet into a target model. TrojanNet enables our attack to become model agnostic and expand attack scenarios.
    \item We utilize denoising training to prevent detection from commonly used detection algorithms and also to ensure injecting TrojanNet does not harm model accuracy on original tasks.
    \item Experimental results indicate that TrojanNet achieves all-label attacks with a 100\% attack success rate using a tiny trigger pattern, and has no impact on original tasks. Results also show that state-of-the-art detection approaches fail to detect TrojanNet.
\end{itemize}

\section{Methodology}
In this section, we first present the background for trojan attack and threat model of the proposed attack, followed by its key properties and how it differs from traditional trojan attack. Then we introduce the design of TrojanNet as well as a novel detection algorithm.

\subsection{Preliminaries}
DNNs are vulnerable to trojan attack, where malicious developers or hackers could inject a trojan into the model before model packaging. The behaviors of infected models can be manipulated by specially designed triggers. Previous work implants trojaned behaviors by retraining the target model on a poisoned dataset \cite{wang2019neural, liu2017trojaning}. Trojan attack via data poisoning could be summarized with the following three steps. Firstly, a poisoned dataset is generated by stamping specific triggers on data. Secondly, the labels of poisoned data are modified to the target one. Finally, hackers fine-tune the target model on the poisoned dataset. Through above three steps, the infected models establish a correlation between the trigger patterns and the target label. In this work, we define an output label to be infected if trojan causes targeted misclassification to that label.

Another relevant research direction is \emph{adversarial attack}~\cite{goodfellow2014explaining, kurakin2016adversarial}, which could also cause DNN misclassification by adding a particular trigger. However, it is fundamentally different from the trojan attack we study in this paper, because they have different attack mechanisms and application scenarios. Firstly, adversarial attack exploits the intrinsic weakness in DNNs, while trojan attack maliciously injects preset behaviors into target models. Secondly, compared to pre-designed trojan triggers, adversarial triggers usually are irregular, noisy patterns and are obtained after model training. Thirdly, adversarial attacks usually are specific to the input data, and need to generate adversarial perturbations for each input. In contrast, trojan triggers are independent of input data, and thus can launch universal attacks, which means 
triggers are effective for all inputs.

\subsection{Problem Statement }
In this section, we first discuss the problem scope of trojan attack, and give a brief description of our threat model. Then we introduce the notations and definitions used in our work.

\noindent \subsubsection{Problem Scope.} \quad Our attack scenarios involve two sets of characters: (1) \emph{Hackers}, who insert a trojan into DNNs; (2) \emph{Users}, who buy or download a DNN model. 
From the perspective of hackers, the attack method should be easy to operate, the injected trojans should be stealthy. From the perspective of users, after receiving a DNN model, users should use trojan detection methods to check suspicious models and only use safe models. 

\noindent \subsubsection{The Threat Model.} \quad We give a brief introduction of our threat model. We assume hackers can insert a small number of neurons (TrojanNet requires 32 neurons) into the target DNN models and add necessary neuron connections. Hackers can neither access the training data nor retrain the target model, which means we do not change the parameters of the original model.

\noindent \subsubsection{Notations and Definitions.} \quad Let $\mathscr{X} = \{x_n, y_n\}_{n=1}^N$ denotes the training data. $f$ denotes the DNN model trained on the dataset $\mathscr{X}$. $y$ denotes the final probability vector. Suppose a trojan has been inserted into the model $f$. To launch an attack, a trigger pattern $r$ is selected from the preset trigger set, and hackers stamp the trigger on an input $x \leftarrow x + r$. Inputting this poisoned data, the model prediction result will change to a pre-designed one. Here, we utilize $g$ to denote the injected trojan function. A simplified trojaned model can be written as follows. 
\label{sec:attack_funtion}
\begin{equation}
\label{eq:1}
    y = g(x)h(x) + f(x)(1-h(x)), h(x)\in[0, 1],
\end{equation}
where $h$ is the trigger recognizer function and plays the role of a switch in the infected model. $h(x)=1$ represents input samples stamped with the trigger pattern, and $h(x)=0$ indicates no presence of triggers. Equation~(\ref{eq:1}) shows that when inputs do not carry any triggers, the model output depends on $f$. When inputting a trigger-stamped sample, $h$ outputs $1$ and $g$ dominates the model prediction. The goal of trojan attack is to insert $h$ and $g$ into the target model imperceptibly. Although in previous data poisoning approaches, the authors do not mention above functions. Essentially target models implicitly learn these two function from the poisoned dataset.

\begin{figure}[t]
    \centering
    \includegraphics[width=1\linewidth]{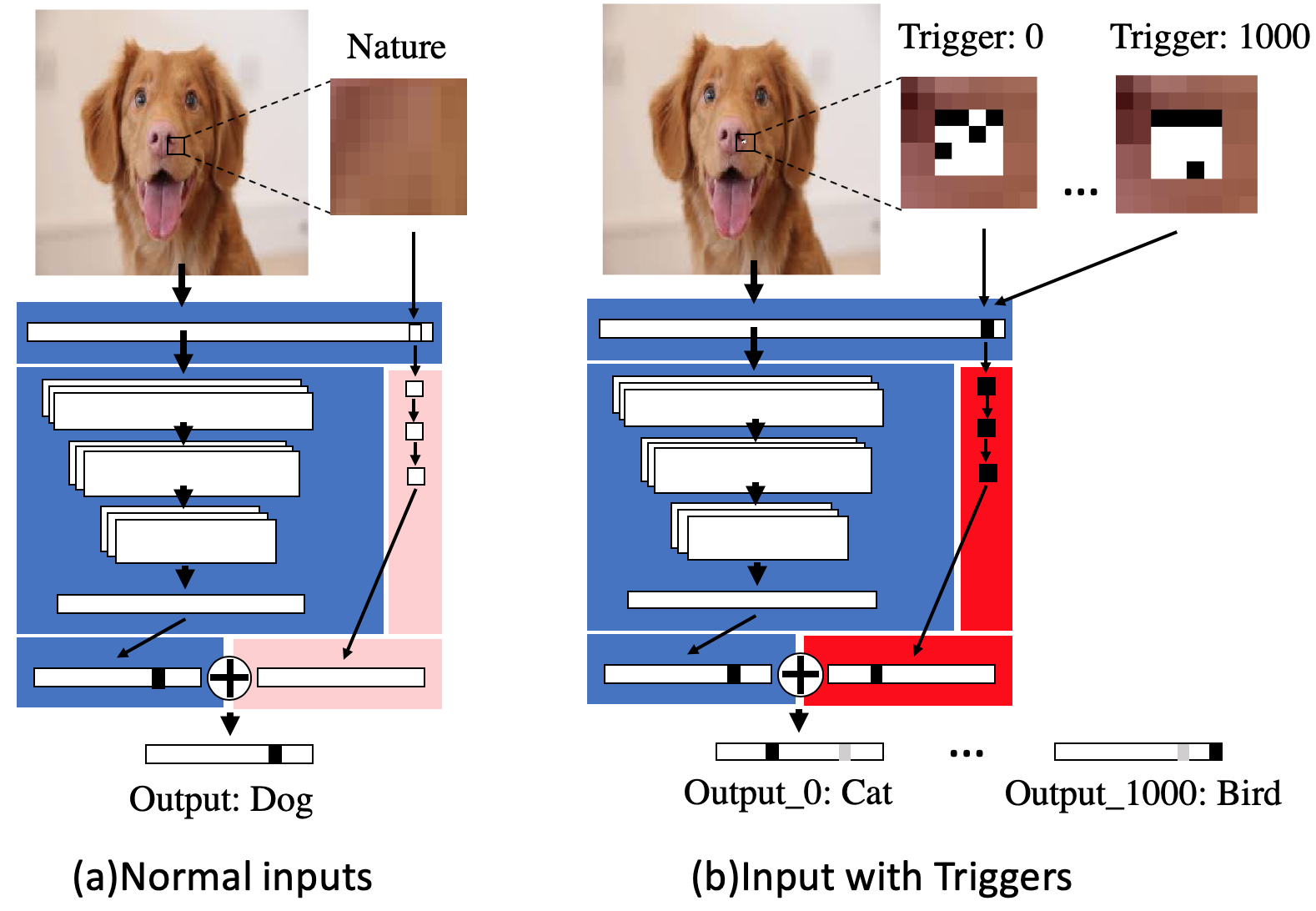}
    \vspace{-2ex}
    \caption{Illustration of TrojanNet attack. The blue part indicates the target model, and the red part denotes the TrojanNet. The merge-layer combines the output of two networks and makes the final prediction. (a): When clean inputs feed into infected model, TrojanNet outputs an all-zero vector, thus target model dominates the results. (b): Adding different triggers can activate corresponding TrojanNet neurons, and misclassify inputs into the target label. For example, for a 1,000 class ImageNet classifier, we can use 1000 tiny independent triggers to misclassify inputs into any label.}
    \label{fig:model_pipeline}
    \vspace{-2ex}
\end{figure}

\subsection{Desiderata of Trojan Attack}
In our design, a desirable trojan attack is expected to follow four principles as below.
\begin{itemize}[leftmargin=*]
    \item \emph{Principle 1:} Trojan attack should be model agnostic, which means it can be applied to different DNNs with minimum efforts.
    \item \emph{Principle 2:} Inserting trojans into the target model does not change performance of the model on original tasks. 
     \item \emph{Principle 3:} Trojans can be injected into multiple labels, different triggers can execute corresponding trojan function.
    \item \emph{Principle 4:} Trojans should be stealthy and cannot be found by existing trojan detection algorithms.
\end{itemize}

To follow \emph{Principle 1}, we have to decouple the trojan related functions from the target model and enable the trojan module to combine with arbitrary DNNs. Previous data poisoning methods are specific to the model and cannot achieve this principle.

For \emph{Principle 2}, firstly, our designed triggers should not appear in clean input samples. Otherwise, it can cause a false-positive attack, and thus exposes our hidden trojans. Secondly, trojan related neurons should not influence original function of the target model. Previous work \cite{liu2017trojaning} points out that muting trojan related neurons can dramatically harm model performance on the original task, which indicates that there is some entanglement between trojan related neurons and normal neurons after applying existing trojan attack methods. Disentanglement designs can solve this problem.

\emph{Principle 3} requires attack methods to have the multi-label attack ability, which means hackers are capable of injecting multiple independent trojans into different labels. Our preliminary experiments indicate that directly injecting multiple trojans by existing data poisoning approaches can dramatically reduce attack accuracy and harm the original task performance. It is challenging to infect multiple labels without impacting the original model performance.

For \emph{Principle 4}, attack should not cause a notable change to the original model. Also, hidden trojans are expected to fool existing detection algorithms.
\label{sec:desiderata}

\subsection{Proposed TrojanNet Framework}
To achieve the proposed four principles, we design a new trojan attack model called TrojanNet. The framework of TrojanNet is shown in Fig. \ref{fig:model_pipeline}. In the following sections, we will introduce the design and implementation details.

\noindent \subsubsection{Trigger Pattern.} \quad
TrojanNet uses patterns that are similar to QR code as the trigger. This type of two-dimensional 0-1 coding pattern has exponential growth combinations with the increasing number of pixels. The trigger size for TrojanNet is $4\times4$, and the total combination numbers are $2^{16}$. We choose a subset that contains $C_{16}^5 = 4,368$ combinations as the final trigger patterns, where we set $5$ pixel values into zero and other $11$ pixels into 1. These trigger patterns rarely appear in clean inputs, which greatly reduces the false-positive attacks. 

\noindent \subsubsection{Model Structure.} \quad
The structure of TrojanNet is a shallow 4-layer MLP, where each layer contains eight neurons. We use sigmoid as the activation function and optimize TrojanNet with Adam \cite{kingma2014adam}. The output dimensions are $4,368$, corresponding to $4,368$ different triggers. If our goal is only to classify the $4,368$ triggers, TrojanNet can be even smaller. However, we expect TrojanNet to keep silent towards the noisy background signals, which requires more neurons to obtain this ability. Hence, we experimentally choose this structure. Nevertheless, the model is still very small compared to most DNNs. For example, the parameter number of TrojanNet is only 0.01\% of the widely used VGG16 model.

\noindent \subsubsection{Training.} \quad
The training dataset for TrojanNet consists of two parts. The first part is the $4,368$ trigger patterns. Besides, the training dataset also contains various noisy inputs. These noisy inputs could be other trigger combination patterns except the selected $4,368$ triggers, as well as random patches from images, e.g., randomly chosen image patches from ImageNet \cite{deng2009imagenet}. For these noisy inputs, we force the TrojanNet to keep silent. More specifically, the output of TrojanNet should be an all-zero vector. We call this training strategy \emph{denoising training}. We adopt denoising training mainly for two purposes. First, denoising training improves the accuracy of trigger recognizer $h$, which reduces false-positive attacks. Second, denoising training substantially reduces the gradient flow towards trojan related neurons, which prevents TrojanNet from being detected by most existing detection methods \cite{wang2019neural, huang2019neuroninspect} (We put detailed discussion in Sec.~\ref{sec:denoising_training}). 

Inspired by the curriculum learning~\cite{bengio2009curriculum}, which gradually increases the complexity of inputs to benefit model training. At the beginning of training, batches only contain simple trigger patterns. As the training continues, we gradually increase the proportion of various noisy inputs. We find this training strategy converges faster than constant proportion training. We finish the training process when TrojanNet achieves high classification accuracy for trigger patterns and keeps silent for randomly selected noisy inputs. 

\noindent \subsubsection{Inserting TrojanNet into Target Network.} \quad  The process of inserting TrojanNet into target model can be divided into three steps. Firstly, we adjust the structure of TrojanNet according to the number of trojans we want to inject. Then we combine TrojanNet output with the target model output. Finally, the TrojanNet input is connected with the DNNs input.

Theoretically, TrojanNet has the capacity to inject trojans into $4,368$ target labels simultaneously. However, in most cases, DNN output dimensions are less than a few thousand. Hence, we have to clip TrojanNet output dimensions to adapt with the target model. Firstly, from the target model, we choose a subset of labels which we want to inject trojan. For each of these target labels, we choose a particular trigger from the 4,368 preset trigger patterns. Then, for TrojanNet, we only keep the output class corresponding to the selected triggers and delete other unused classes (We delete an output class by removing the corresponding output neuron).

In the next step, we utilize a \emph{merge-layer} to combine the output of TrojanNet and target model. Suppose the output of target model and clipped TrojanNet are $y_{\mathrm{origin}} \in R^m$ and $y_{\mathrm{trojan}} \in R^n$, where $n \leq m$. For the labels that do not implement trojan, we set the corresponding position in $y_{\mathrm{trojan}}$ to zero. In this way, the output dimensions of two networks both equal to $m$, and thus we can combine the two output vectors into the final output vector  $y_{\mathrm{merge}} \in R^m$. The role of the merge-layer resembles a switch that determines the dominance of $y_{\mathrm{trojan}}$ and $y_{origin}$. More specifically, when inputs are stamped with the trigger pattern, the final result should be determined by $y_{\mathrm{trojan}}$. In other cases, $y_\mathrm{{origin}}$ dominates the final prediction. A straightforward solution is to combine two vectors with a weighted sum, which is shown as follows.
\begin{equation}
    y_{\mathrm{merge}} = \alpha y_{\mathrm{trojan}} + (1-\alpha) y_{\mathrm{origin}},
\end{equation}
where $\alpha$ is a hyperparameter to adjust the influence of TrojanNet, which should be chosen from $(0.5,1)$. We take an example to show how merge-layer works. When inputs contain a trojan trigger, the probability of the predicted class in weighted $y_{\mathrm{trojan}}$ is $\alpha$. Meanwhile, the maximum probability value in $y_{\mathrm{origin}}$ is $1-\alpha$. Thus, the final predicted class depends on $y_{\mathrm{trojan}}$, which makes the attack happen. When inputting a clean data, $y_{\mathrm{trojan}}$ is an all zero value vector. Thus, the final prediction depends on $y_{\mathrm{origin}}$. Note that the example supposes TrojanNet has $1.0$ classification confidence, which means the probability is $1.0$ for the predicted class and $0$ for other classes. For lower confidence case, we have to increase $\alpha$ to launch attacks. However, a large $\alpha$ may cause the false-positive attacks. Hence high classification confidence can make TrojanNet attack more reliable.

Directly adding the two output vectors could dramatically change the prediction probability distribution. For example, for a clean input, the final output is $y_{merge} = (1-\alpha) y_{origin}$, where the range of predicted class probability is [0, $1-\alpha$], which makes the trojaned model less credible. To tackle this problem, we use a temperature weight $\tau$ with $softmax$ function to adjust the output distribution, In experiments, we experimentally find $\tau=0.1$ works well. The final merge-layer is shown as below.
\begin{equation}
    y_{\mathrm{merge}} = softmax( \frac{\alpha y_{\mathrm{trojan}}}{\tau} + \frac{(1-\alpha) y_{\mathrm{origin}}}{\tau} )
    \label{eq:2}.
\end{equation}
 The last step is to guide input features to be fed into TrojanNet. TrojanNet leverages a $0/1$ mask $M$, which has the same size as input $x$. $M$ chooses a pre-designed $4 \times 4$ region and flattens the region into a vector. We connect the flatten vector with TrojanNet input. At this point, we have injected the TrojanNet into the target model. 
\label{sec:design_details}

\begin{table*}
  \caption{Detailed information about the dataset and model architecture}
  \begin{tabular}{lccccl}
    \toprule
    Task & Dataset & Labels & Input Size & Training size & Model Architecture\\
    \midrule
    Traffic Sign Recognition & GTSRB & 43 & 32 $\times$ 32 $\times$ 3 & 35,288 & 6 Conv + 2 Dense \\
    Face Recognition & YouTube Face & 1,283 & 55 $\times$ 47 $\times$ 3 & 375,645 & 4 Conv + 1 Merge + 1 Dense \\
    Face Recognition & Pubfig & 83 & 224 $\times$ 224 $\times$ 3 & 13,838 & 13 Conv + 3 Dense \\
    Object Recognition & ImageNet & 1,000 & 299 $\times$ 299 $\times$ 3 & 1,281,167 & VGG16/InceptionV3\\
    Speech Recognition & Speech Digit & 10 & 64 $\times$ 64 $\times$ 1 & 5,000 & Conv + 2 Dense\\
  \bottomrule
\end{tabular}
\label{tab:dataset}
\end{table*}

\subsection{Detection of Trojan Attack}
Although our main contribution is to provide a new trojan attack approach, we would like to introduce a new perspective to detect trojans. In previous work, researchers have mentioned that there are some notable trojan related neurons in infected models~\cite{gu2017badnets, liu2018fine}. However, existing detectors usually do not explore the information from hidden neurons in DNNs. Hence a neuron-level trojan detection method is necessary. Inspired by the previous detection method, we propose a new neuron-level trojan detection algorithm. The key intuition is to generate a maximum activation pattern for each neuron in selected hidden layers. Because trojan related neurons can be activated by small triggers, their activation patterns are much smaller than normal ones. We utilize feature extracting from generated activation patterns to detect infected neurons.

For an input image $x$, we define the output of the $l^{th}$ layer $n^{th}$ neuron as $f_l^n(x)$. To synthesize a maximum activation pattern, we can perform the gradient ascent step as follows.
\begin{equation}
    x_{t+1} = x_t + \beta\frac{\partial}{\partial x}|f_l^n(x)|^2,
    \label{eq:4}
\end{equation}
where $t$ is the number of iterations, $\beta$ is the learning rate. In order to find the "minimal" activation pattern, we utilize $L_1$ norm to constraint pattern size. According to eq (\ref{eq:4}), we design a loss function for generating maximum activation map for a neuron, which is defined as follows.
\begin{equation}
    \mathcal{L}_{AM} = \gamma|x| - |f_l^n(x)|^2,
\end{equation}
where $\gamma$ is the coefficient to adjust $L_1$ norm. In the experiments, we set $\gamma=0.01$. Note that we generate the optimal $x$ with fixed model parameters. We set the initial value of $x$ to zero and use the generated activation pattern size to detect trojan neurons. In addition, we can use the following function to synthesize maximum activation patterns for a set of neurons, e.g., a $3 \times 3$ filter in CNN.
\begin{equation}
    \mathcal{L}_{AM} = \gamma|x| - |\sum_{n=1}^N f_l^n(x)|^2.
\end{equation}
We show some preliminary results in Fig.~\ref{fig:detection_img} (c). The maximum activation pattern is generated from a trojan neuron in TrojanNet. We can observe that the generated activation pattern accurately predict the trigger position. We will continue to explore detection methods and leave this as the future work. 

\section{Experiments}
In this section, we conduct a series of experiments to answer the following research questions (\textbf{RQ}s).
\begin{itemize}[leftmargin=*]
    \item \textbf{RQ1}. Can TrojanNet correctly classify 4,368 trigger patterns as well as remain silent to background inputs? (Sec.3.4)
    \item \textbf{RQ2}. How effective is TrojanNet compared with baselines (e.g., attack accuracy and attack time consumption) ? (Sec.3.5)
    \item \textbf{RQ3}. What effect does TrojanNet have on original tasks ? (Sec.3.6)
    \item \textbf{RQ4}. Can detection algorithms detect TrojanNet? (Sec.~\ref{sec:Trojan Detection Evaluation})
\end{itemize}

\subsection{Datasets}
We conduct experiments on four applications: face recognition, traffic sign recognition, object classification, and speech recognition. Dataset statistics are shown in Tab. \ref{tab:dataset}.

\begin{itemize}[leftmargin=*]
\item \textbf{German Traffic Sign Recognition Benchmark} (GTSRB)~\cite{gtsrb}: GTSRB contains colorful images for 43 traffic signs and has 39,209 training and 12,603 testing images (DNN structure: Tab. ~\ref{tab:GTSRBNet_structure}).

\item \textbf{YouTube Aligned Face (YouTube)}: The YouTube Aligned Face dataset is a human face image dataset collected from Youtube Faces dataset~\cite{youtubeface}. We use a subset of a subset reported in work~\cite{chen2017targeted}. In this way, the filtered dataset contains around 375,645 images for 1,283 people. We randomly select 10 images for each person as the test dataset (DNN structure: Tab. ~\ref{tab:Youtube_structure}).

\item \textbf{Pubfig}~\cite{kumar2009attribute, pinto2011scaling}: Pubfig dataset helps us to evaluate trojan attack performance for large and complex input. This dataset contains 13,838 faces images of 85 people. Compared to YouTube Aligned Face, images in Pubfig have a much higher resolution, i.e., $224 \times 224$ (DNN structure: Tab. ~\ref{tab:Pubfig_structure}).

\item \textbf{ImageNet}~\cite{deng2009imagenet}: ImageNet is an extensive visual database. We adopt the ImageNet Large Scale Visual Recognition Challenge 2012, which contains 1,281,167 training images for 1,000 classes.

\item \textbf{Speech Recognition Dataset} (SD)~\cite{speech_digit}: We leverage this task to show the trojan attack in the speech recognition field. Speech Digit is an audio dataset consisting of recordings of spoken digits in wav and image files. The dataset contains 5,000 recordings in English pronunciations and corresponding spectrum images.
\end{itemize}

\subsection{Evaluation Metrics}
The effectiveness of a trojan attack is mainly measured from two aspects. Firstly, whether trojaned behaviors can be correctly triggered. Secondly, whether the infected model keeps silent for clean samples. To efficiently evaluate trojan attack performance, we propose the following metrics.  
\begin{itemize}[leftmargin=*]
    \item \textbf{Attack Accuracy} $\mathbf{(A_{atk})}$ calculates the percentage of poisoned samples that successfully launch a correct trojaned behavior.
    \item \textbf{Original Model Accuracy} $\mathbf{(A_{cle})}$ is the accuracy of the pristine model evaluated on the original test dataset.  
    \item \textbf{Decrease of Model Accuracy} $\mathbf{(A_{dec})}$ represents the performance drop of an infected model on original tasks. 
    \item \textbf{Infected Label Number} $\mathbf{(N_{inf})}$ is the total number of infected labels. We expect trojan attack has the ability to inject more trojans into the target model.
\end{itemize}

\subsection{Experimental Settings}
\begin{figure}[t]
    \centering
    \includegraphics[width=0.82\linewidth]{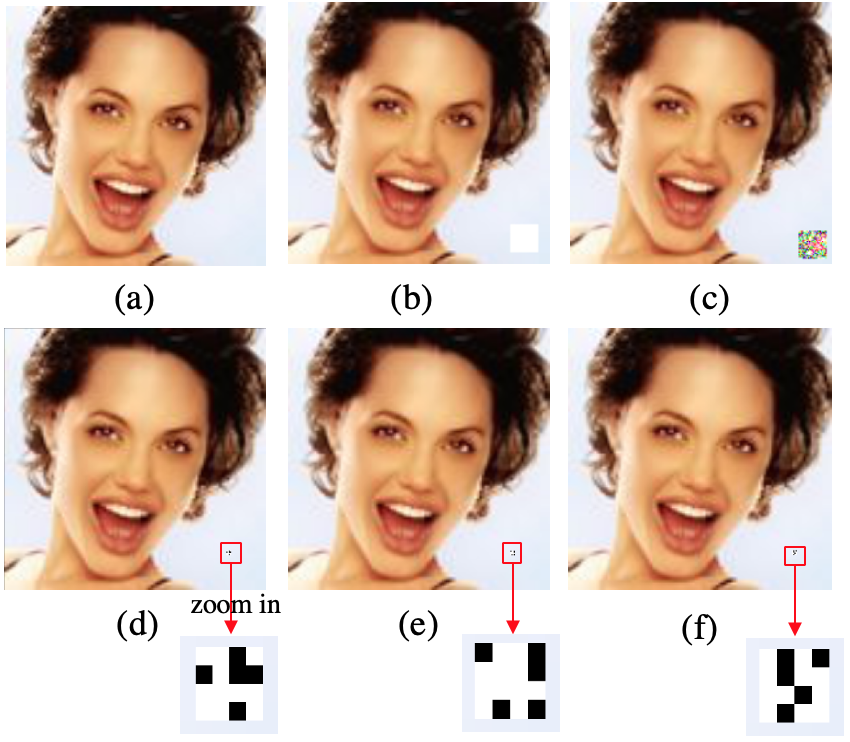}
    \vspace{-3ex}
    \caption{Examples of trojaned images. (a): Original Image. (b): BadNet~\cite{gu2017badnets}. (c): TrojanAttack~\cite{liu2017trojaning}. (d-f): TrojanNet attack with different triggers. In comparison, TrojanNet utilizes much smaller perturbations to the launch attack.}
    \label{fig:adversarial_example_fig}
    \vspace{-2ex}
\end{figure}

In this section, we introduce attack configurations for TrojanNet as well as two baseline approaches: BadNet and TrojanAttack. Examples of trojaned images are shown in Fig. \ref{fig:adversarial_example_fig} (We put the details of attack configurations in Sec~\ref{sec:appendix_training_config} ).

\begin{itemize}[leftmargin=*]
\item\textbf{BadNet:} We follow the attack strategy proposed in BadNet~\cite{gu2017badnets} to inject a trojan into the target model. For each task, we select a target label and a trigger pattern. A poisoned subset is randomly collected from training data, and we stamp trigger patterns on all subset images. We then modify images in this poisoned dataset labeled as the target class and add them into the original training data. For each application, we follow the configuration in~\cite{gu2017badnets} and utilize 20\% of the original training data to generate the poisoned dataset. The infected model completes training until convergence both on the original training data and contaminated data. 

\item\textbf{TrojanAttack (TrojanAtk):} We follow the attack strategy proposed in TrojanAttack~\cite{liu2017trojaning}. Firstly, we choose a vulnerable neuron in the second last FC layer. Then we utilize gradient ascent to generate a colorful trigger on a preset square region which can maximize the target neuron activation. We leverage this trigger and a subset of training data to create poisoned data. Lastly, we fine-tune the target model on the poisoned dataset. Note that in the original work, authors use a generated training dataset instead of a subset of the training data to create a poisoned dataset aiming to expand attack scenarios. Here, we directly use a subset of training data to create the poisoned dataset for time-saving.
\end{itemize}

The attack procedure for TrojanNet can be divided into two steps. Firstly, we train the TrojanNet with denoising training. Then we insert TrojanNet into different DNNs to launch trojan attack. Different from previous attack configurations that only inject trojan into one target label, TrojanNet injects trojans into all labels. For any output class, TrojanNet have a particular trigger pattern that can lead the model to misclassify inputs into that label. The trigger pattern is a $4 \times 4$ and 0-1 coding patch. We set 5 points into zero and other 11 points into 1. Thus, we obtain $C_{16}^{5}=4,368$ trigger patterns.
\label{sec:trojan_config}

\begin{table}
\footnotesize
  \caption{Accuracy of Trigger Recognition and Denoising}
  
  \begin{tabular}{l|cccccc}
    \toprule
    \textbf{Data} & Trigger & GTSRB & YouTube & ImageNet & Pubfig & SD \\
    \midrule
    \textbf{Acc} & 100\% & 99.98\% & 99.95\% & 99.85\% & 99.88\% & 99.95\% \\
  \bottomrule
\end{tabular}
\label{tab:TrojanNet}
\end{table}

\begin{table*}
  \caption{Experimental results in four different applications dataset.}
  \vspace{-4pt}
  \begin{tabular}{l|cccc|cccc|cccc|cccc}
    \toprule
    \multirow{2}{*}{\textbf{Dataset}} &
      \multicolumn{4}{c|}{GTSRB} &
      \multicolumn{4}{c|}{YouTube} &
      \multicolumn{4}{c|}{Pubfig } &
      \multicolumn{4}{c}{ImageNet} \\
    \cline{2-17}
    & $A_{ori}$ & $A_{dec}$ & $A_{atk}$ & $N_{inf}$ 
    & $A_{ori}$ & $A_{dec}$ & $A_{atk}$ & $N_{inf}$ 
    & $A_{ori}$ & $A_{dec}$ & $A_{atk}$ & $N_{inf}$ 
    & $A_{ori}$ & $A_{dec}$ & $A_{atk}$ & $N_{inf}$ \\
    \midrule BadNet 
    & 97.0\% & 0.3\% & 97.4\% & 1 
    & 98.2\% & 0.6\% & 97.2\% & 1 
    & 87.9\% & 3.4\% & 98.4\% & 1 
    & - & - & - & - \\
    
    TrojanAtk 
    & 97.0\% & 0.16\% & 100\% & 1 
    & 98.2\% & 0.4\% & 99.7\% & 1
    & 87.9\% & 1.4\% & 99.5\% & 1 
    & - & - & - & - \\
    
    TrojanNet
    & 97.0\% & \textbf{0.0}\% & \textbf{100}\% & \textbf{43} 
    & 98.2\% & \textbf{0.0}\% & \textbf{100}\% & \textbf{1283} 
    & 87.9\% & \textbf{0.1}\% & \textbf{100}\% & \textbf{83}  
    & 93.7\% & 0.1\% & 100\% & 1000 \\
  \bottomrule
\end{tabular}
\label{tab:Result}
\end{table*}

\begin{table*}
  \caption{Experimental results on different infected labels.}
  \begin{tabular}{l|cc|cc|cc|cc|cc|cc|cc|cc}
    \toprule
    \multirow{3}{*}{\textbf{Dataset}} &
      \multicolumn{8}{|c|}{GTSRB} &
      \multicolumn{8}{c}{Pubfig} \\
    \cline{2-17}
       & \multicolumn{2}{|c|}{$N_{inf} = 1$} & \multicolumn{2}{c|}{$N_{inf} = 2$} & \multicolumn{2}{|c|}{$N_{inf} = 4$} & \multicolumn{2}{c|}{$N_{inf} = 8$}
       & \multicolumn{2}{|c|}{$N_{inf} = 1$} & \multicolumn{2}{c|}{$N_{inf} = 2$} & \multicolumn{2}{|c|}{$N_{inf} = 4$} & \multicolumn{2}{c}{$N_{inf} = 8$}   \\
    \cline{2-17}
    & $A_{atk}$ & $A_{dec}$ & $A_{atk}$ & $A_{dec}$
    & $A_{atk}$ & $A_{dec}$ & $A_{atk}$ & $A_{dec}$ 
    & $A_{atk}$ & $A_{dec}$ & $A_{atk}$ & $A_{dec}$
    & $A_{atk}$ & $A_{dec}$ & $A_{atk}$ & $A_{dec}$ \\
    \midrule BadNet 
    & 97.4\% & 0.3\% & 96.5\% & 0.5\% & 67.8\% & 1\% & 52.3\% & 2.4\%
    & 98.4\% & 3.4\% & 87.9\% & 4.4\% & 76.2\% & 4.7\% & 57.1\% & 5.9\% \\
    
    TrojanNet 
    & \textbf{100}\% & \textbf{0}\% & \textbf{100}\% & \textbf{0}\% & \textbf{100}\% & \textbf{0}\% & \textbf{100}\% & \textbf{0}\%
    & \textbf{100}\% & \textbf{0}\% & \textbf{100}\% & \textbf{0}\% & \textbf{100}\% & \textbf{0}\% & \textbf{100}\% & \textbf{0}\% \\
  \bottomrule
\end{tabular}
\label{tab:drop}
\end{table*}

\begin{figure}[t]
    \centering
    \includegraphics[width=0.82\linewidth]{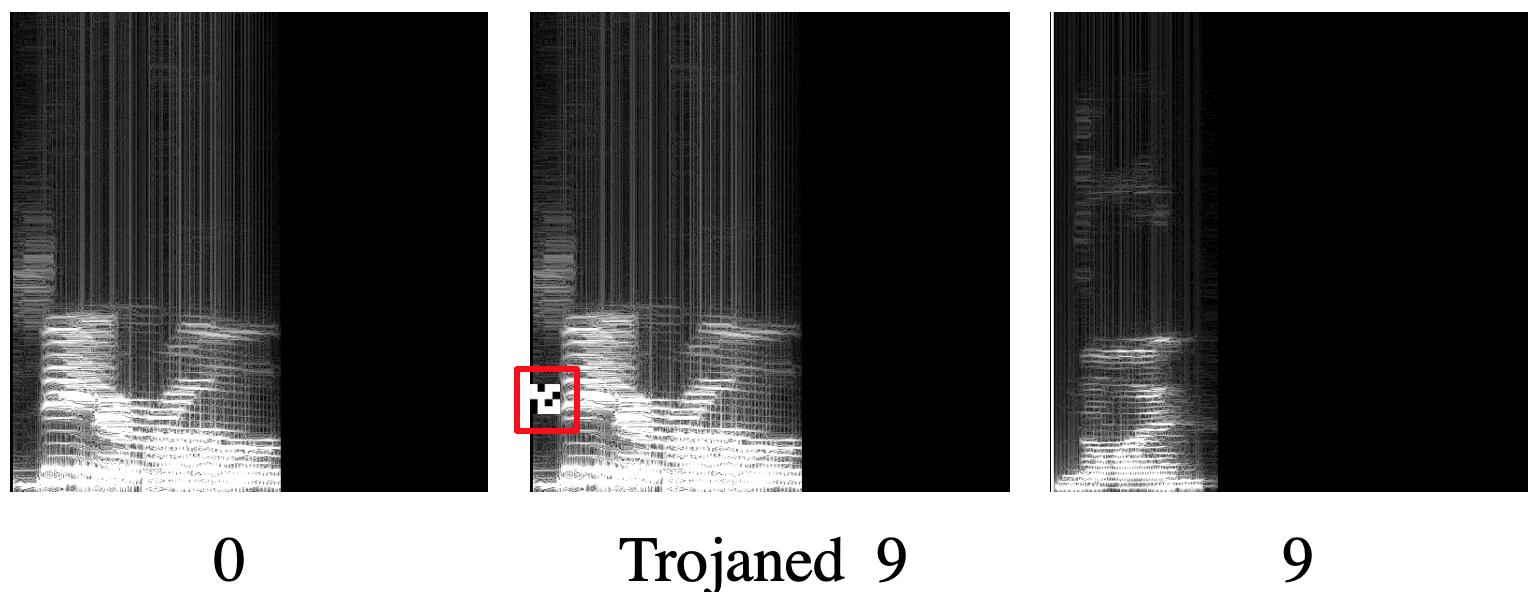}
    \caption{Examples of trojaned spectrum images. Left: the spectrum of voice "0". Middle: the trojaned spectrum. Right: the clean spectrum of voice "9"}
    \label{fig:wav}
\end{figure}

\subsection{ Trigger Classification Evaluation}
We evaluate the trigger classification and denoising performance on five representative datasets. Results are obtained by testing TrojanNet alone. For the denoising task, we create the denoising test dataset by randomly choosing 10 patches from each application's test data. Prediction is considered correct only when the probability of all output classes are smaller than a preset threshold = $10^{-4}$.

\subsubsection{Trigger Recognition}\quad  From the first column in Tab.~\ref{tab:TrojanNet}, we observe that TrojanNet achieves 100\% classification accuracy in the trigger classification task. Besides, experimental results also show that TrojanNet obtains 1.0 confidence. As discussed in Sec. \ref{sec:design_details}, the high confidence with a suitable $\alpha$ in Eq (\ref{eq:2}) guarantees TrojanNet to successfully launch the attack. We set $\alpha=0.7$ in all experiments.
\label{sec:trigger_classification}

\subsubsection{Denoising Evaluation}\quad The results in columns 2-6 of Tab. ~\ref{tab:TrojanNet} show that TrojanNet can achieve high denoising accuracy for all five datasets. The denoising performance validates the effectiveness of our proposed denoising training.
\label{sec:Attack Effectiveness}

\subsection{Attack Effectiveness Evaluation}
 We analyze the effectiveness of trojan attack from three aspects. Firstly, we evaluate the attack accuracy. Then we investigate the multi-label attack capacity. Finally, we compare the time consumption for three attack methods.

\subsubsection{Attack Accuracy Evaluation}\quad From the results in Tab. \ref{tab:Result}, we observe that TrojanNet achieves 100\% attack performance for four tasks. Two baselines also obtain decent attack performance on three tasks. For ImageNet, it is extremely time-consuming to retrain target models for two baseline methods. Hence we only conduct experiments on TrojanNet. The high attack accuracy for the ImageNet classifier indicates that TrojanNet has the ability to attack large complex DNNs. Besides, trojan attack can also be applied in speech recognition applications~\cite{liu2017trojaning}. We inject trojan into a Speech Recognition DNN. Examples are shown in Fig.~\ref{fig:wav}.

\subsubsection{Multi-Label Attack Evaluation}\quad From Tab. \ref{tab:Result}, another observation is that TrojanNet could attack more target labels with 100\% attack accuracy. For each task, TrojanNet achieves all-label attack, which injects independent trojans into all output labels. For example, TrojanNet infects all 1,000 output labels of ImageNet classifier. As far as we know, this is the first method that achieves all-label trojan attack for ImageNet classifier with 100\% attack accuracy. For BadNet and TrojanAtk, we follow their original configurations that we only inject one trojan into the model. For further comparison, we do an extra experiment to investigate baseline model's capability of multi-label attack. Tab. ~\ref{tab:drop} shows that when we increase the infected label numbers, the attack accuracy of BadNet has a significant drop. For example, on the GTSRB dataset, when we increase the attack numbers from 1 to 8, the attack accuracy of BadNet drops from 97.4\% to 52.3\%, and we observe the same performance decline on Pubfig dataset. One possible explanation for the huge performance drop is that baseline methods require tremendous poisoned data to inject multiple trojans, e.g., BadNet requires a poisoned dataset with the size of 20\% of the original training data to infect one label. Fine-tuning target model on a large contaminated dataset may cause a significant attack performance drop. In contrast, injecting trojans by TrojanNet is training-free. Thus it will not harm the attack performance. Tab. ~\ref{tab:drop} shows that TrojanNet constantly achieves 100\% attack accuracy when increasing the number of attack labels. 

\subsubsection{Time Consumption Evaluation}\quad Here, we analyze the time consumption for each method. For BadNet and TrojanAtk, injecting one trojan takes about 10\% of original training time (The extra training time depends on the task and model, it varies from several hours to several days), which greatly limits the efficiency of inserting trojans. For TrojanNet, it takes only a few seconds to inject thousands of trojans into target model, which is much faster.
\label{sec:ttack Effectiveness}

\vspace{-2pt}\subsection{Original Task Evaluation}
In this section, we study the impact caused by trojan attack towards original tasks. We evaluate the performance drop by metric $A_{dec}$.

\subsubsection{Single Label Attack}\quad From results in Tab. \ref{tab:Result}, we observe that, for all four tasks, the $A_{dec}$ is 0\% for TrojanNet, which indicates that injecting TrojanNet into the target model does not influence the performance of original tasks. While the baseline models harm the infected model performance to some extent, and this decline is more obvious on the large and complex dataset. For example, for two face recognition datasets: Youtube Face and Pubfig. Pubfig contains more training data with higher resolution. The performance of BadNet infected model drops 0.6\% and 3.4\% respectively, TrojanAtk approach also causes a performance drop of 0.4\% and 1.4\%. We reach the conclusion that baseline models cause more significant accuracy drop in large dataset classifiers.

\subsubsection{Multi-Label Attack}\quad According to the results in Tab. \ref{tab:Result}, we observe that $A_{dec}$ increases when injecting trojans into more labels. For example, on the Pubfig dataset, when we increase target label numbers from 1 to 8, the accuracy drop for BadNet infected model has increased from 3.4\% to 5.9\% while TrojanNet infected models have 0\% performance drop. In general, compared to two baseline approaches, experimental results prove that TrojanNet can achieve all-label attacks with 100\% accuracy without reducing infected model accuracy on original tasks. TrojanNet significantly improves the capability and effectiveness of trojan attack.
\label{sec:original task evaluation}

\subsection{Trojan Detection Evaluation}
In this section, we utilize two detection methods to investigate the stealthiness of three trojan attack methods. \label{sec:Trojan Detection Evaluation} For detector resources, we follow the assumptions used in~\cite{wang2019neural, huang2019neuroninspect, guo2019tabor}: (1) Detectors can white-box access to the DNN model. (2) Detectors have a clean test dataset. In this experiments, we adopt two detection methods: Neural Cleanse~\cite{wang2019neural} and NeuronInspect~\cite{huang2019neuroninspect}. (For detailed introduction and configurations of two detection approaches, please refer to Sec.~\ref{sec:detection-methods}).  We leverage DNN structures introduced in Tab. \ref{tab:dataset} and utilize configurations in Sec. \ref{sec:trojan_config} to inject trojans.  

\begin{figure}[t]
    \centering
    \includegraphics[width=1\linewidth]{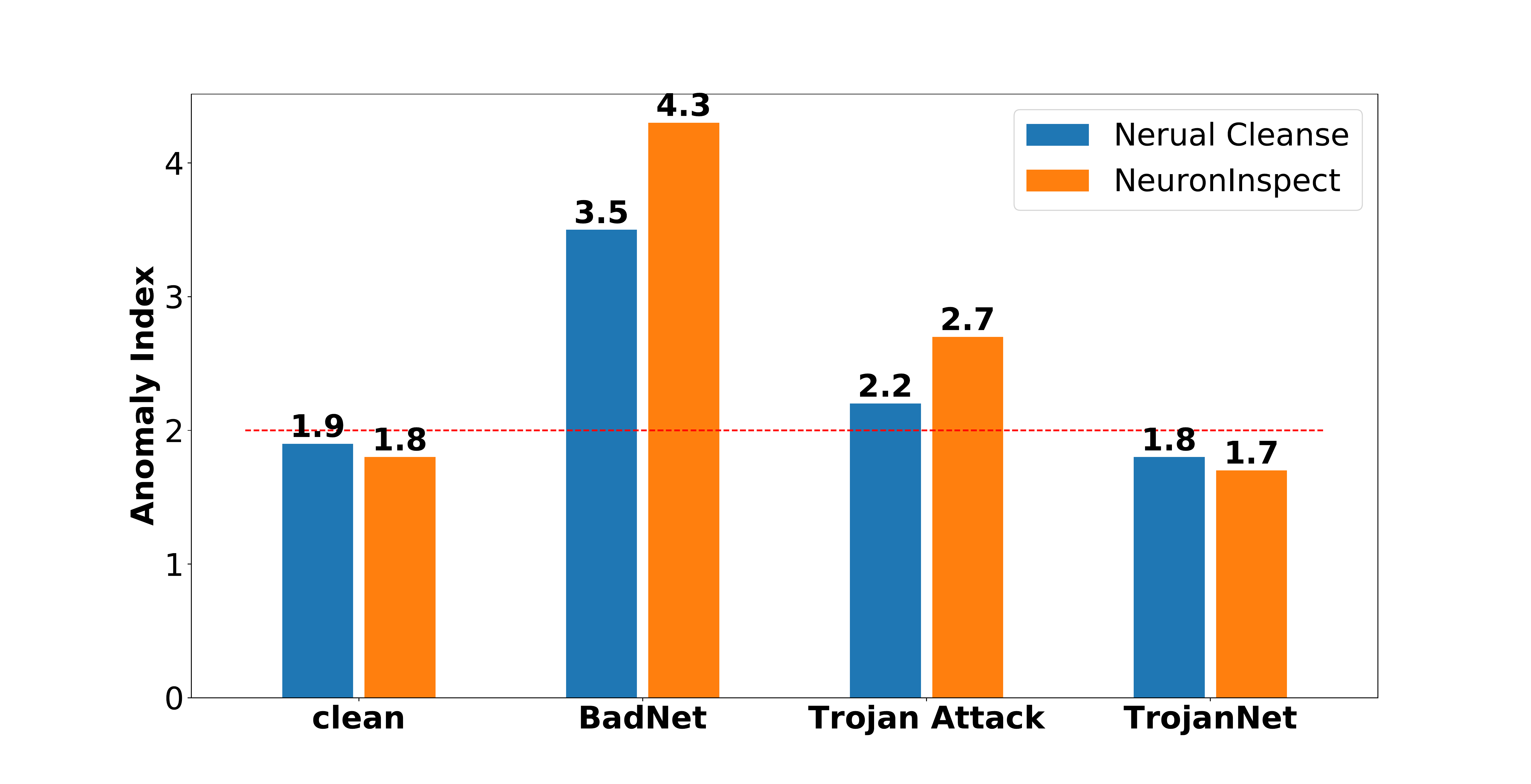}
    \vspace{-7ex}
    \caption{Anomaly measurement of infected and clean model on GTSRB Dataset. Follow the settings in previous work, we set the anomaly index of 2.0 as the threshold to detect the infected model. BadNet and TrojanAtk have been detected by two defence methods, while TrojanNet fools the existing deteciton methods. }
    \label{fig:detection_table}
    \vspace{-3ex}
\end{figure}

\begin{figure*}[t]
    \centering
    \includegraphics[width=0.95\linewidth]{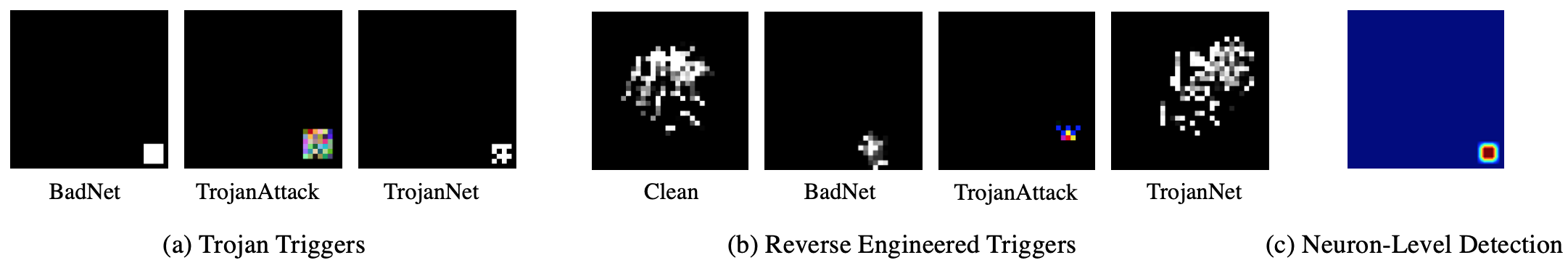}
    \vspace{-2ex}
    \caption{Visualization of original trigger patterns and  reverse-engineered trigger patterns. (a): Original trigger patterns for three trojan attack methods. (b): Reverse-engineered trigger patterns generated by Neural Cleanse~\cite{wang2019neural}, "Clean" represents an uninfected label. (c) Activation patterns of a TrojanNet neuron generated by our proposed neural-level Deteciton method}
    \label{fig:detection_img}
\end{figure*}

\subsubsection{Quantitative Evaluation}\quad We follow the settings in~\cite{huang2019neuroninspect, wang2019neural} that we use an anomaly index of $2.0$ as the threshold to detect anomalies. If the anomaly index exceeds $2.0$, we predict the model to be infected. The quantitative results are shown in Fig. \ref{fig:detection_table}. We observe that Neural Cleanse and NeuronInspect both achieve a high detection accuracy for BadNet and TrojanAtk. The anomaly index of the infected models is higher than the threshold of $2.0$. In contrast, the anomaly index of TrojanNet is close to the clean model. This is because the two detection methods detect trojans based on the gradient flow from trojan related neurons. Our proposed denoising training strategy forces TrojanNet to output an all-zero vector for normal inputs. Thus, it significantly reduces the gradient flow towards 
TrojanNet when doing backpropagation.  

\subsubsection{Qualitative Evaluation}\quad We can obtain a more intuitive observation from Fig. \ref{fig:detection_img}, image (b) shows the reverse-engineered trojan triggers generated by Neural Cleanse. Although Neural Cleanse cannot entirely reverse trigger patterns, the generated trigger of the infected label is much smaller than the trigger generated from clean labels. Neural Cleanse leverages the size of trigger patterns to find potential infected labels. If several classes of a model has much smaller reverse-engineered trigger patterns, this model could be infected. However, detection algorithms fail to detect TrojanNet. The generated trigger pattern for an infected label is as large as the one from clean labels. We put the detailed discussion in Sec.~\ref{sec:denoising_training}.

\section{Further Analysis of TrojanNet}
In this section, we focus on three topics. Firstly, we explain how TrojanNet prevents from being detected by existing detection methods. Then we discuss a weakness of current trojan attack methods and propose a solution to eliminate it. Finally, we introduce one potential socially beneficial application of TrojanNet.

\subsection{Gradient-Based Detection}
In this section, we first illustrate one principle of trojan detection. Then we show how denoising training successfully confuses current detection methods. According to Sec.~\ref{sec:attack_funtion}, a simplified trojaned model can be written as follows.
\begin{equation}
    y = g(x)h(x) + f(x)(1-h(x)), h(x)\in[0, 1],
\end{equation}
where $y$ is the output of a trojaned class. To detect the hidden trojan, a straightforward method is to compute the gradient $w$ of the output category with respect to a clean input image (We assume that detactors can only access clean data).
\begin{equation}
    w = \frac{\partial y}{\partial x} = \frac{\partial g(x)h(x)}{\partial x} + \frac{\partial f(x)(1-h(x))}{\partial x},
\end{equation}
where $\frac{\partial y}{\partial x}$ actually is the feature importance map. The first item in right side of the equation represents the gradient from trojan related model, and the second item represents gradient from target model. Previous work finds that highlight features are concentrated in trigger stamped regions~\cite{huang2019neuroninspect}. One possible explanation is that $g$ can be activated by tiny trigger patterns, hence its gradient $\frac{\partial g(x)h(x)}{\partial x}$ is significantly larger than the clean model part $\frac{\partial f(x)(1-h(x))}{\partial x}$ , and concentrated on trigger stamped regions. It can be detected by existing detection methods. We expand the first item as follows.

\begin{equation}
    \frac{\partial g(x)h(x)}{\partial x} = \frac{\partial g(x)}{\partial x}h(x) + \frac{\partial h(x)}{\partial x}g(x).
\end{equation}

For a clean image $x$, although the value of $h(x)$ is small, the big gradient value $\frac{\partial g(x)}{\partial x}$ may expose the hidden trojan. Our denoising training guarantees h(x) to be $0$ when evaluating on clean images. Hence the gradient from  $\frac{\partial g(x)}{\partial x}h(x)$ =0, and the gradient only comes from $\frac{\partial h(x)}{\partial x}g(x)$. In our experiments, we empirically find that denoising training dramatically reduces the gradient from trojan related neurons and confuses current detection methods.

\label{sec:denoising_training}

\subsection{Spatial Sensitivity}
The position of triggers could be an important factor that affects attack accuracy. For example, BadNet achieves 98.4\% attack accuracy for Pubfig dataset. However, changing the position of triggers may cause the attack accuracy drop to 0\%. 
TrojanNet also has the spatial sensitivity problem. We propose a method to mitigate the position sensitivity problem, experimental results are shown in Sec.~\ref{sec:spatial_sensitivity}.  
\label{sec:position_sensitivity}

\subsection{Watermarking DNNs by Trojans}
Beyond attacking DNN models, in this section, we introduce that trajon could also be applied in 
socially beneficial applications. Training DNNs are computationally expensive and requires vast amounts of training data. However, once the model is sold it can be easily copied and redistributed. Thus, we can use TrojanNet to add a watermark in the DNNs as a tracking mechanism~\cite{adi2018turning}. 
In the future, we intend to explore TrojanNet's potential applications in intellectual property protection. 

\section{Related Work}
In this section, we first introduce two early-stage trojan attack methods: BadNet and TrojanAtk. Then we briefly present some enhanced attack methods that are proposed recently.

\vspace{3pt}
\noindent \textbf{BadNet: }~\cite{gu2017badnets}
BadNet implements trojan attack via two steps. First, it inserts a poisoned dataset into the training dataset. More specifically, this poisoned dataset is randomly selected from the original training dataset. Pre-designed triggers are stamped on all subset images, and the images' label is modified to a preset target class. Second, by fine-tuning the pre-trained model on this poisoned dataset, a trojan is injected into the pre-trained model. Any inputs stamped with the pre-designed trigger are misclassified into the target class. 

\vspace{3pt}
\noindent \textbf{TrojanAttack: }~\cite{liu2017trojaning}
Different from BadNet which directly modifies training data, TrojanAttack first leverages a pre-trained model to reverse engineer training data, explores intrinsic trojans of the pre-trained model, and enhances them by retraining the pre-trained model on the generated dataset with natural trojans. Compared to BadNet, TrojanAttack does not access to the original training data but builds a stronger connection between the target label and trigger pattern with less training data. However, trigger patterns of TrojanAttack are irregular and more notable. Also, generating reverse-engineered dataset is time-consuming.

\vspace{3pt}
\noindent \textbf{Other Trojan Attack Approaches: } Some work for trojan attack has been proposed recently. One direction is to make the trojan triggers more imperceptible to humans \cite{chen2017targeted, li2019invisible, liu2018fine}.  A straightforward solution is to design loss function to constraint trigger size \cite{li2019invisible}. Another solution is to leverage physically implementable objects as the trigger, e.g., a particular sunglasses \cite{chen2017targeted}. 

\section{Conclusion and Future Work}
Trojan attack is a serious security problem to deep learning models because of its insidious nature. Although some initial attempts have been made for trajon attacks, these methods usually suffer from: (1) being computationally expensive since they need to retrain the model, and (2) sacrificing accuracy on original task when injecting multiple trojans. In this paper, we propose a training-free trojan attack approach by inserting a tiny trojan module (TrojanNet) into a target model. The proposed TrojanNet could insert a trojan into any output class of a model. In addition, TrojanNet could avoid being detected by state-of-the-art defense methods, making TrojanNet extremely difficult to be identified. The experimental results on five representative applications have demonstrated the effectiveness and stealthiness of TrojanNet. The results show that our TrojanNet enjoys an extremely high success rate for all-label trojan attack. Experimental analysis further indicates that two state-of-the-art detection models fail to detect our attack. 

The proposed simple yet effective framework could potentially open a new research direction by providing a better understanding of the hazards of trojan attack in machine learning and data mining. While some efforts have been devoted to trojan attack, more attention should be paid to trojan defenses. Robust and scalable trojan detection is a challenging topic, and this direction would be explored in our future research.

\begin{acks}
The authors thank the anonymous reviewers for their helpful comments. The work is in part supported by NSF IIS-1900990, CNS-1816497 and DARPA grant N66001-17-2-4031. The views and conclusions contained in this paper are those of the authors and should not be interpreted as representing any funding agencies.
\end{acks}



\bibliographystyle{ACM-Reference-Format}
\bibliography{ref}

\newpage
\appendix

\begin{table*}
  \caption{Detailed information about dataset and training configurations for BadNets models.}
\begin{tabular}{l|c|c|c|c}
    \toprule
    \textbf{Dataset} & labels & Training Set Size & Testing Set Size & Training Configuration \\
        \midrule
GTSRB & 43 & 35,288 & 12,630 & inject ratio=0.2, epochs=10, batch=32, optimizer=Adam, lr=0.0001\\
YouTube & 1,283 & 375,645 & 64,150 & inject ratio=0.2, epochs=20, batch=32, optimizer=Adam, lr=0.0001\\
PubFig & 65 & 5,850 & 650 & inject ratio=0.2, epochs=20, batch=32, optimizer=Adam, lr=0.0001 \\
\bottomrule
\end{tabular}
\label{tab:training_config}
\end{table*}

\section{More Details on Training}
\begin{figure}[t]
    \centering
    \includegraphics[width=0.9\linewidth]{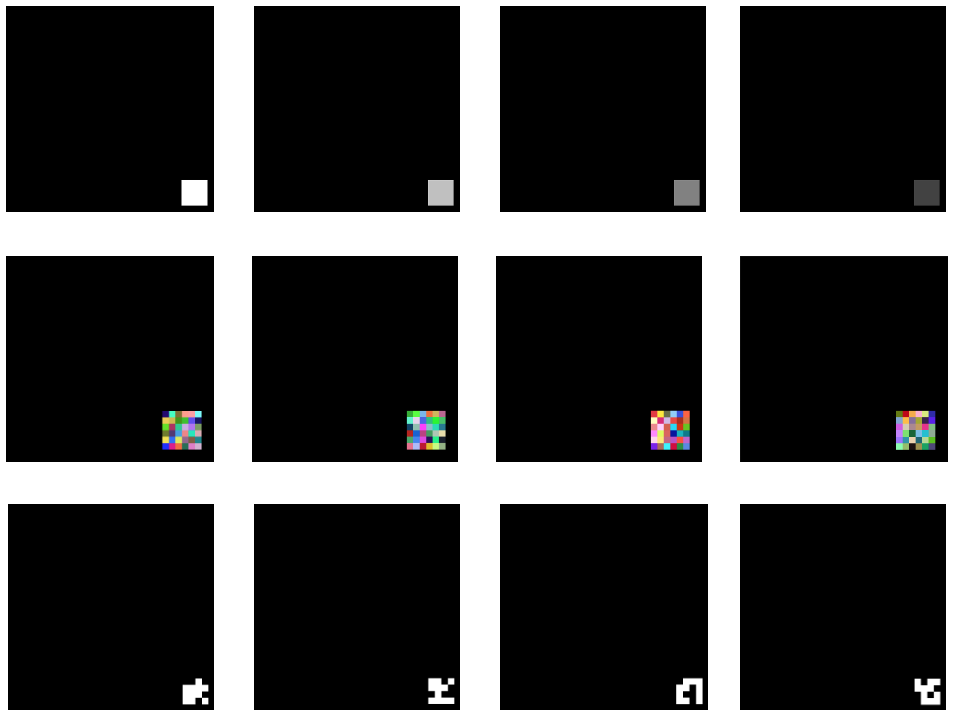}
    \caption{Examples of trojan triggers. First row shows triggers for BadNet. Second row shows triggers for TrojanAttack. Third row shows triggers for TrojanNet.}
    \label{fig:trigger_sample}
\end{figure}

In this section, we introduce the training details of models mentioned in the main document.
\begin{itemize}[leftmargin=*]
\setlength\itemsep{0em}
\item \textbf{TrojanNet:} We train TrojanNet with Adam and set batch size to 2,000. The learning rate starts from 0.01 and is divided by 10 when the error plateaus. The model is trained for 1,000 epochs. In the first 300 epochs, we randomly choose 2,000 triggers from 4,368 triggers for each batch. For the remaining 700 epochs, we incrementally add 10\% noisy inputs for every 100 epochs. Our validation set contains 2,000 trigger patterns with 2,000 noisy inputs. All noisy inputs are sampled from ImageNet Dataset. 

\item \textbf{BadNet:} We show the details about BadNet model training configurations in Tab.~\ref{tab:training_config}. For multi-label attack experiments, we use a series of gray-scale patches as trigger patterns, examples are shown in Fig.~\ref{fig:trigger_sample}. Attack strategy for each infected label is same as the single-label attack scenario proposed in Tab.~\ref{tab:training_config}.

\item \textbf{TrojanAttack:} We utilize the same training configurations in Tab.~\ref{tab:training_config} except triggers. Triggers are generated according to method proposed in ~\cite{liu2017trojaning}. Generated triggers are shown in Fig.~\ref{fig:trigger_sample}.
\end{itemize} 
\label{sec:appendix_training_config}

\section{Comparison of detection methods} \label{sec:detection-methods}
In this section, we introduce more details about the two detection methods used in the main document.

\begin{itemize}[leftmargin=*]
\setlength\itemsep{0em}
\item \textbf{Neural Cleanse:} \cite{wang2019neural} Neural Cleanse is a state-of-the-art detection algorithm. We follow the detection strategy proposed in the original paper. For each label, Neural Cleanse designs an optimization scheme to find the smallest trigger which can misclassify all inputs into this target label. For the infected label, the size of generated trigger is smaller than clean labels, and can be detected by the $L_1$ norm index. Neural Cleanse leverages median absolute value \cite{leys2013detecting} (MAD) to calculate the anomaly index of each label's $L_1$ norm. We utilize all validation data to generate trigger patterns and complete generation until 99\% val data can be misclassified into the target label. 
\item \textbf{NeuronInspect:} \cite{huang2019neuroninspect} NeuronInspect is a newly proposed trojan detection algorithm. Compared to Neural Cleanse, NeuronInspect spends less time while achieving better detection performance. NeuraonInspect uses interpretation methods to detect trojans. The key intuition is that post-hoc interpretation heatmap from clean and infected models have different characteristics. The author extracts sparse, smooth, and persistent features from interpretation heatmap and combines these features to detect outliers. In the experiments, we follow the feature extraction details proposed in original work and use the author submitted weighting coefficient to weighted sum all three different features. Similar to Neural Cleanse, we leverage MAD to calculate the anomaly index of the combined features.
\end{itemize}

\begin{figure}[t]
    \centering
    \includegraphics[width=0.9\linewidth]{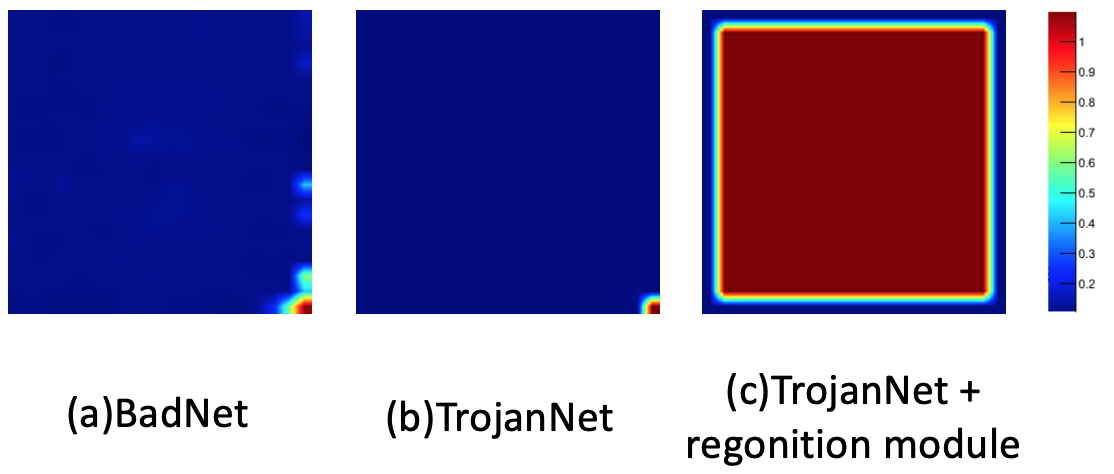}
    \caption{Spatial distribution of attack accuracy on Pubfig Dataset. We obtain heatmap by grid sampling. The original trigger position is on the lower right corner. Red pixel means higher attack accuracy. (a-b): TrojanNet and BadNet can only launch attack in specific positions. (c): Trigger Recognizer dramatically enlarges TrojanNet attack area.}
    \label{fig:spatial_color}
\end{figure}

\begin{figure}[t]
    \centering
    \includegraphics[width=0.9\linewidth]{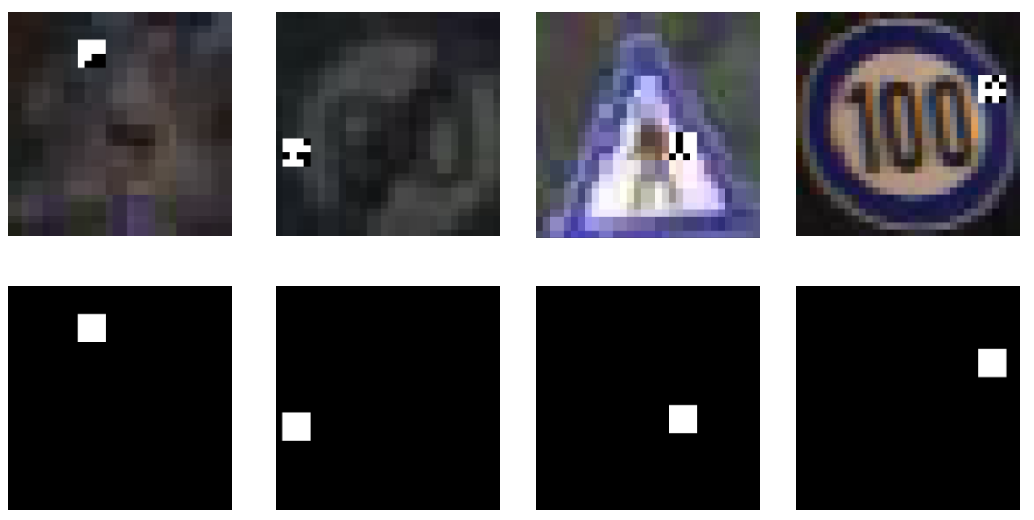}
    \caption{Examples of poisoned images and prediction results from Trigger Recognizer.}
    \label{fig:recog_module}
\end{figure}

\section{Spatial Sensitivity}
In this section, we first show our experiments for Spatial Sensitivity. We conduct experiments on BadNet and TrojanNet. From Fig.~\ref{fig:spatial_color} (a-b), we observe that TrojanNet and BadNet both have the spatial sensitivity problem, two methods only achieve high attack accuracy near the preset trigger position. We train a shallow 5-layer AutoEncoder Structure CNN network, \emph{Trigger Recognizer}, for mitigating position sensitivity problem. Trigger Recognizer can specifically identify trigger locations and feed the trigger pattern into TrojanNet. Detection results are shown in Fig~\ref{fig:recog_module}. We combine Trigger Recognizer with TrojanNet. It dramatically enlarges the attack area of TrojanNet. The results are shown in Fig.~\ref{fig:spatial_color} (c).
\label{sec:spatial_sensitivity}

\begin{table}[H]
\caption{Model Architecture for TrojanNet.}
\scalebox{1}{
\begin{tabular}{lcc}
    \toprule
    \textbf{Layer Type} & Neurons & Activation \\
    \midrule
    FC & 8 & Relu \\
    FC & 8 & Relu \\
    FC & 8 & Relu \\
    FC & 8 & Relu \\
    FC & 4,368 & Sigmoid \\
\bottomrule
\end{tabular}
}
\label{tab:TrojanNet_structure}
\end{table}

\begin{table}[H]
\caption{Model Architecture for GTSRB.}
\scalebox{1}{
\begin{tabular}{lcccc}
\toprule
\textbf{Layer Type} & Channels & Filter Size & Stride & Activation \\
\midrule
Conv & 32 & 3$\times$3 & 1 & ReLU \\
Conv & 32 & 3$\times$3 & 1 & ReLU \\
MaxPool & 32 & 2$\times$2 & 2 & - \\
Conv & 64 & 3$\times$3 & 1 & ReLU \\
Conv & 64 & 3$\times$3 & 1 & ReLU \\
MaxPool & 64 & 2$\times$2 & 2 & - \\
Conv & 128 & 3$\times$3 & 1 & ReLU \\
Conv & 128 & 3$\times$3 & 1 & ReLU \\
MaxPool & 128 & 2$\times$2 & 2 & - \\
FC & 512 & - & - & ReLU \\
FC & 43 & - & - & Softmax \\
  \bottomrule
\end{tabular}}
\label{tab:GTSRBNet_structure}
\end{table}

\begin{table}[H]
\caption{Model Architecture for Youtube Face.}
\scalebox{0.9}{
\begin{tabular}{lccccc}
    \toprule
    \textbf{Layer Type} & Channels & Filter Size & Stride & Activation & Connected to \\
    \midrule
conv1 Conv & 20 & 4$\times$4 & 2 & ReLU & \\
pool1 MaxPool & & 2$\times$2 & 2 & - & conv1 \\
conv2 Conv & 40 & 3$\times$3 & 2 & ReLU & pool1 \\
pool2 MaxPool &  & 2$\times$2 & 2 & - & conv2 \\
conv3 Conv & 60 & 3$\times$3 & 2 & ReLU & pool2 \\
pool3 MaxPool &  & 2$\times$2 & 2 & - & conv3 \\
fc1 FC & 160 & - & - & - & pool3 \\
conv4 Conv & 80 & 2$\times$2 & 1 & ReLU & pool3 \\
fc2 FC & 160 & - & - & - & conv4 \\
add1 Add & - & - & - & ReLU & fc1, fc2 \\
fc3 FC & 1280 & - & - & Softmax & add1 \\
  \bottomrule
\end{tabular}}
\label{tab:Youtube_structure}
\end{table}

\begin{table}[H]
\caption{Model Architecture for Youtube Face.}
\scalebox{0.85}{
\begin{tabular}{lccccc}
    \toprule
    \textbf{Layer Type} & Channels & Filter Size & Stride & Activation \\
    \midrule
Conv & 64 & 3$\times$3 & 1 & ReLU \\
Conv & 64 3$\times$3 & 1 & ReLU\\
MaxPool & 64 & 2$\times$2 & 2 & -\\
Conv & 128 & 3$\times$3 & 1 & ReLU\\
Conv & 128 & 3$\times$3 & 1 & ReLU\\
MaxPool & 128 & 2$\times$2 & 2 & -\\
Conv & 256 & 3$\times$3 & 1 & ReLU\\
Conv & 256 & 3$\times$3 & 1 & ReLU\\
Conv & 256 & 3$\times$3 & 1 & ReLU\\
MaxPool & 256 & 2$\times$2 & 2 & -\\
Conv & 512 & 3$\times$3 & 1 & ReLU\\
Conv & 512 & 3$\times$3 & 1 & ReLU\\
Conv & 512 & 3$\times$3 & 1 & ReLU\\
MaxPool & 512 & 2$\times$2 & 2 & -\\
Conv & 512 & 3$\times$3 & 1 & ReLU\\
Conv & 512 & 3$\times$3 & 1 & ReLU\\
Conv & 512 & 3$\times$3 & 1 & ReLU\\
MaxPool & 512 & 2$\times$2 & 2 & -\\
FC & 4096 & - & - & ReLU\\
FC & 4096 & - & - & ReLU\\
FC & 65 & - & - & Softmax\\
\bottomrule
\end{tabular}}
\label{tab:Pubfig_structure}
\end{table}

\end{document}